\documentclass[superscriptaddress,amsmath,amssymb,reprint,nobibnotes,footinbib,prb]{revtex4-1}

\usepackage{graphicx}
\usepackage[colorlinks=true,linkcolor=blue,citecolor=blue,urlcolor=blue]{hyperref}
\usepackage{multirow}

\begin{document}

\title{Spontaneous surface flux pattern in chiral $p$-wave superconductors  - revisited}
\author{Sarah B. Etter}
\affiliation{Institute for Theoretical Physics, ETH Zurich, 8093 Zurich, Switzerland}
\author{Adrien Bouhon}
\affiliation{Department of Physics and Astronomy, Uppsala University, Box 516, SE-751 20 Uppsala, Sweden}
\author{Manfred Sigrist}
\affiliation{Institute for Theoretical Physics, ETH Zurich, 8093 Zurich, Switzerland}
\date{\today}

\begin{abstract}
In chiral $p$-wave superconductors, magnetic flux patterns may appear spontaneously when translational symmetry is broken such as at surfaces, domain walls, or impurities. However, in the candidate material Sr$_2$RuO$_4$ no direct signs of such magnetic fields have been detected experimentally. In this paper, the flux pattern at the edge of a disk-shaped sample is examined using the phenomenological Ginzburg Landau approach. The detailed shape of the flux pattern, including self-screening, is computed numerically for different surface types by systematically scanning a range of boundary conditions. Moreover, specific features of the electronic structure are included qualitatively through the coefficients in the Ginzburg Landau functional. Both the shape and the magnitude of the flux pattern are found to be highly sensitive to all considered parameters. 
In conclusion, such spontaneous magnetic flux patterns are not a universal feature of chiral $p$-wave superconductors.
\end{abstract}

\maketitle

\section{Introduction}\label{sec:intro}

Superconductors with chiral $p$-wave pairing have been studied extensively because the broken time reversal symmetry and their topological nature lead to a variety of interesting phenomena\cite{mackenzie:2003,sigrist:2005a,kallin:2012,maeno:2012}. In particular, topology-related chiral edge states give rise to quasiparticle currents along the surface, transporting both energy and charge\cite{sigrist:2000}. While these two are obviously related and emerge from the same underlying physics, their features are very different. The topologically protected quasiparticle (energy) current, connected to the charge neutral Majorana zero mode, is described straightforwardly \cite{kallin:2016} and would be experimentally accessible, for example, through the quantized thermal Hall conductivity\cite{sumiyoshi:2013,imai:2016}. The charge edge current and the resulting spontaneous magnetic flux pattern near the surface, on the other hand, is a more obscure feature both quantitively and qualitatively as charge is not a conserved property of the Bogolyubov quasiparticles\cite{huang:2015}. Thus, direction and magnitude of supercurrents at the edge subtly depend on microscopic details of the bulk electronic states and scattering properties of the surface. 

The superconductor Sr$_2$RuO$_4$ is the best-known candidate for chiral $p$-wave pairing. Several experiments point to broken time reversal symmetry\cite{luke:1998,xia:2006} and spin-triplet pairing\cite{ishida:1998,liu:2003, anwar:2016}. However, other experimental results inconsistent with this interpretation keep the debate about the pairing symmetry ongoing\cite{kallin:2012}. A prominent challenge are the null results in the search for the edge currents in both scanning Hall bar\cite{kirtley:2007} and scanning SQUID\cite{hicks:2010,curran:2014} experiments, as well as in cantilever magnetometry\cite{jang:2011}. A review of possible experiments with quantitative estimates is given by Kwon, Yakovenko and Sengupta\cite{kwon:2003}. There have been various theoretical proposals related to this issue, via both microscopic and phenomenological arguments. 
As a `reference scenario', Matsumoto and Sigrist assumed specular scattering at a planar surface and an isotropic Fermi surface\cite{matsumoto:1999}. Concerning the surface type, Ashby and Kallin considered rough and pair-breaking surfaces in a phenomenological Ginzburg Landau (GL) approach\cite{ashby:2009}, while Lederer \emph{et al.} considered a metallic surface layer in a lattice Bogoliubov-de Gennes (BdG) approach supplemented with phenomenological arguments\cite{lederer:2014}. They both also analyzed the impact of changing the GL coefficients.
Concerning the real-space geometry, Huang and Yip studied a small-sized disk geometry for both specular and diffusive scattering in an external magnetic field\cite{huang:2012}. Sauls discussed the influence of retroreflection using a quasiclassical approach\cite{sauls:2011}. Concerning the electronic structure, Becerra \emph{et al.} considered an anisotropy of the Fermi surface\cite{becerra:2016} in a GL approach, and for an applied field. 
Bouhon and Sigrist\cite{bouhon:2014} studied a lattice BdG model for specular scattering and found a non-trivial dependence of the edge current direction on the surface orientation in combination with band structure effects beyond an isotropic Fermi surface. Finally, other approaches include multi-band theories\cite{raghu:2010}, higher Chern numbers\cite{scaffidi:2015} or higher angular momentum triplet pairing\cite{huang:2014,suzuki:2016}. 
These diverse treatments all show a reduction of the magnetic field with respect to the `reference scenario'.
Importantly, any explanation for the absence of magnetic fields at the edge has to be reconciled with the magnetic signatures observed in $\mu$SR experiments\cite{luke:1998}.

In this paper, the different propositions regarding the surface magnetic flux pattern of a chiral $p$-wave superconductor are combined and extended systematically for a comprehensive analysis based on the phenomenological Ginzburg Landau (GL) approach. The detailed shape of the surface magnetic flux pattern is computed numerically, including self-screening, for a disk geometry such that all surface orientations can be observed simultaneously. The boundary conditions are scanned through a range of surface properties beyond specular scattering. In addition, details of the electronic structure as described in Ref.~[\onlinecite{bouhon:2014}], away from the isotropic limit, are reflected in the choice of the GL coefficients. The corresponding full GL model is constructed in Sec.~\ref{sec:model}, while the parameter ranges which are systematically scanned are discussed in Sec.~\ref{sec:scan}. The results are analyzed in detail in Sec.~\ref{sec:res}. First, the effect of the surface scattering types and of the anisotropy are examined separately. Next, combining both features, the full set of flux patterns is presented and analyzed. The main conclusion of this study is the fact that the structure and magnitude of the magnetic flux pattern near the disk edge are very sensitive to the parameters considered, i.e. the surface scattering properties as well as the electronic anisotropy of the superconductor.
Interestingly, in a certain parameter range the total magnetic flux of the disk vanishes despite being finite locally near the edge. In addition, the flux pattern at impurities is briefly addressed in Sec.~\ref{sec:imp}. 

\section{Model, parameters and method}\label{sec:model}

The Ginzburg Landau (GL) formalism provides an ideal framework to systematically study the surface magnetic flux pattern of a chiral $p$-wave superconductor for a range of material and surface properties which can be covered by varying only a small number of parameters in the GL free energy. 
The gap function of the spin-triplet chiral $p$-wave state can conveniently be represented using the standard d-vector notation
\begin{equation}
\boldsymbol{d}_{\pm} (\boldsymbol{k}) = \eta_0 (k_x \pm i k_y) \boldsymbol{\hat{z}},
\end{equation}
for which $|\Delta_{\boldsymbol{k}}|^2 = |\boldsymbol{d}(\boldsymbol{k})|^2$ is the square of the quasiparticle energy gap. This state is two-fold degenerate, indicated by $ \pm $ denoting the positive and negative chirality (broken time-reversal symmetry), and the orientation $ \boldsymbol{d} \parallel \boldsymbol{\hat{z}} $ specifies the in-plane equal-spin pairing (ESP) spin configuration.
The chiral $p$-wave gap function resides in the vector space of the two-dimensional irreducible odd-parity representation $ E_u $ of the tetragonal point group $ D_{4h} $ (crystal lattice symmetry) with the basis functions having the symmetry $ \{ k_x , k_y \} $.  
Introducing the complex two-component order parameter $ \boldsymbol{\eta} = (\eta_x , \eta_y) $, the general in-plane ESP state is expressed as $ {\boldsymbol{d}}(\boldsymbol{k})=(\eta_x k_x+\eta_yk_y)\boldsymbol{\hat{z}}$, where both components can depend on spatial coordinates independently, while the bulk chiral $p$-wave state has the form $ \boldsymbol{\eta}= \eta_b (1, \pm i ) $. 


\subsection{Ginzburg Landau functional}

The Ginzburg Landau free energy functional for the bulk of a chiral $p$-wave superconductor in a material with tetragonal crystal symmetry is given by\cite{sigrist:1991}
\begin{align}\label{eq:bulk}
&\mathcal{F}[\eta_x,\eta_y,\boldsymbol{A}]=\int\mathrm{d}^3r \Bigg[a\left(T-T_c\right)|\boldsymbol{\eta}|^2\\
&\quad+b_1|\boldsymbol{\eta}|^4+\frac{b_2}{2}\left(\eta_x^{*2}\eta_y^2+\eta_x^2\eta_y^{*2}\right)+b_3|\eta_x|^2|\eta_y|^2\nonumber\\
&\quad+K_1\left(|D_x\eta_x|^2+|D_y\eta_y|^2\right)+K_2\left(|D_x\eta_y|^2+|D_y\eta_x|^2\right)\nonumber\\
&\quad+\Big\{K_3(D_x\eta_x)^*(D_y\eta_y)+K_4(D_x\eta_y)^*(D_y\eta_x)+\mathrm{c.c}\Big\}\nonumber\\
&\quad+K_5\left(|D_z\eta_x|^2+|D_z\eta_y|^2\right)+\frac{\left(\nabla\times\boldsymbol{A}\right)^2}{8\pi}\Bigg],\nonumber
\end{align}
with the critical temperature $T_c$ and the covariant derivative $\boldsymbol{D}=\nabla-i\gamma\boldsymbol{A}$, where $\gamma=2e/(\hbar c)=2\pi/\Phi_0$ with the flux quantum $\Phi_0$, and the vector potential $\boldsymbol{A}$ with the magnetic field $\boldsymbol{B}=\nabla\times\boldsymbol{A}$.

The expansion coefficients $a$, $\{b_i\}$ and $\{K_i\}$ for the second order, the forth order and the gradient terms, respectively, depend on material properties and can to some extent be determined by experiments, primarily the ratios between these sets. For example, they enter expressions for the specific heat jump at the superconducting transition, the London penetration depth, or the coherence length connected with the upper critical field. 
Alternatively, the coefficients can be derived from microscopic models, as outlined in Appendix~\ref{app:para}, which in particular helps to determine the ratios within the sets of coefficients $\{b_i\}$ and $\{K_i\}$. Often, coefficients from weak-coupling theories based on an isotropic Fermi surface are used for convenience\cite{machida:1985}. However, effects of Fermi surface anisotropy can be implemented straightforwardly in a quasiclassical framework\cite{agterberg:1998}. Such details of the electronic structure can indeed affect properties of the superconductor qualitatively. An instructive and for our purpose relevant example is the ratio $K_1/K_2$ which is crucial for the behavior of the order parameter near the surface and for the edge currents. It was noticed\cite{bouhon:2014} that this ratio varies strongly with the band filling in a square-lattice tight-binding model, from $K_1/K_2=3$ for an isotropic (cylindrical) Fermi surface to values $K_1/K_2 < 1$ for situations beyond the lowest-order anisotropy found in quasiclassical models. Tuning between these values, a reversal of the current direction can occur for certain surface orientations.

In the following, the ratio $K_1/K_2$ is one of the important parameters. While varying  $K_1/K_2$ we fix $ K = K_1 + K_2 =\mathrm{const}$ and, at the same time, keep  the other coefficients corresponding to the standard isotropic weak-coupling ratios (see Appendix~\ref{app:para}), such that $b_1=3/8\,b$, $b_2=-b_3=b/4$ and $K_3=K_4=K/4$, for simplicity. The bulk order parameter $ \boldsymbol{\eta}_{\rm b} = \eta_{\rm b}(T) (1, \pm i) $ is then obtained by minimizing the GL free energy,
\begin{align}\label{eq:coef:eta0}
|\eta_{\rm b}(T)|^2=\frac{-a(T-T_c)}{4b_1-b_2+b_3}=\frac{-a(T-T_c)}{b}.
\end{align}
The in-plane coherence length $\xi(T)$ can be derived from the GL equation as,
\begin{align}\label{eq:coef:xi}
\xi(T)^2=\frac{K_1+K_2}{-2a(T-T_c)}=\frac{K}{-2a(T-T_c)}.
\end{align}
With this choice the basic length scale of the order parameter remains unchanged while scanning $K_1/K_2$ for a constant $K$. Finally, the London penetration depth is given by
\begin{align}\label{eq:coef:lam}
\lambda_L(T)^{-2}=8\pi\gamma^2(K_1+K_2) |\eta_{\rm b} (T)|^2 = 8\pi\gamma^2K|\eta_{\rm b}|^2 ,
\end{align}
which neither depends on $K_1/K_2$.

These definitions naturally lead to a dimensionless formulation useful for numerical treatment. The temperature is given in units of $ T_c $, lengths in units of $ \xi_0 = \xi(0) $, the order parameter in units of $ \eta_0 = \eta_{\rm b} (0)$, and the vector potential in units of $1/(\gamma\xi_0)$ such that the magnetic field is expressed in units of $H_{c2} (0)$. What remains to be fixed is the GL parameter $ \kappa = \lambda / \xi $ which is chosen to be 2.6 in accordance with measurements\cite{maeno:2012} in Sr$_2$RuO$_4$. Note that we refer here to values of quantities extrapolated to $ T=0$ within the GL approach.

\subsection{Surface effects}

The GL approach provides a particularly simple way to include a variety of surface properties through general boundary terms supplementing the bulk free energy Eq.~(\ref{eq:bulk}). For the chiral $p$-wave superconductor they are given by\cite{sigrist:1991}
\begin{align}\label{eq:surf}
&\mathcal{F}_{\mathrm{surf}}[\eta_x,\eta_y]=\int_{{\rm surf}}\mathrm{d}^2r\Big[\left(g_1\left(n_x^2+n_y^2\right)+g_2n_z^2\right)|\boldsymbol{\eta}|^2\\
&+g_3\left(n_x^2-n_y^2\right)\left(|\eta_x|^2-|\eta_y|^2\right)+g_4n_xn_y\left(\eta_x^{*}\eta_y+\eta_x\eta_y^{*}\right)\Big],\nonumber
\end{align}
with the integral running over the surface of the superconductor and with the normal $\boldsymbol{n}$ pointing outwards. The coefficients $\{g_i\}$ again depend on material properties and also on the surface type. They may in general be spatially dependent. For our study, however, identical conditions are assumed along the whole surface. In the absence of an external field as considered here, the boundary conditions include $\left.B_z\right|_{{\rm surf}} = \left.(\boldsymbol{\nabla} \times \boldsymbol{A})_z \right|_{\rm surf}=0$ for $ \boldsymbol{n} \perp \boldsymbol{\hat{z}} $.

The variation of the free energy (including bulk and surface terms) with respect to each order parameter component leads to the corresponding boundary conditions. 
As an example, for a planar surface at $ \boldsymbol{r}= (x_0,y,z) $ with $\boldsymbol{n}=\boldsymbol{\hat x}$, for which $\boldsymbol{\eta}=\boldsymbol\eta(x)$ and $\boldsymbol{A}=A_y(x)\boldsymbol{\hat y}$ with the gauge $ \boldsymbol{\nabla} \cdot \boldsymbol A = 0 $, the resulting boundary conditions are
\begin{subequations}\label{eq:fullbc}
\begin{align}
\left[K_1\partial_x\eta_x-i\gamma A_yK_3\eta_y+\left(g_1+g_3\right)\eta_x\right]_{x_0}&=0\label{eq:fullbc_x}\\
\left[K_2\partial_x\eta_y-i\gamma A_yK_3\eta_x+\left(g_1-g_3\right)\eta_y\right]_{x_0}&=0.
\end{align}
\end{subequations}

The definition of the extrapolation length $l$ of the order parameter at the surface can be extended to the two-component case straightforwardly as\cite{degennes:1999}
\begin{align}\label{eq:expol}
\left.\frac{|\nabla_{\boldsymbol{n}}\eta_i|}{|\eta_i|}\right|_{{\rm surf}}=\frac{1}{l_i},
\end{align}
where for the above example $ i = (x,y) $. Ignoring the vector potential, i.e. neglecting self-screening, as for example in Refs.~[\onlinecite{ashby:2009,lederer:2014}], the boundary conditions from Eq.~(\ref{eq:fullbc}) with $K_i>0$ and $g_1\geq g_3\geq 0$ (see below) lead to
\begin{align}
\label{eq:extrapol:noA}
l_x=\frac{K_1}{\left(g_1+g_3\right)},\quad l_y=\frac{K_2}{\left(g_1-g_3\right)}.
\end{align}
In a general treatment, the extrapolation lengths according to Eq.~(\ref{eq:expol}) can always be extracted a-posteriori from the resulting shape of the order parameter components.

\subsection{Disk geometry}

\begin{figure}[t]
\includegraphics[width=0.9\columnwidth]{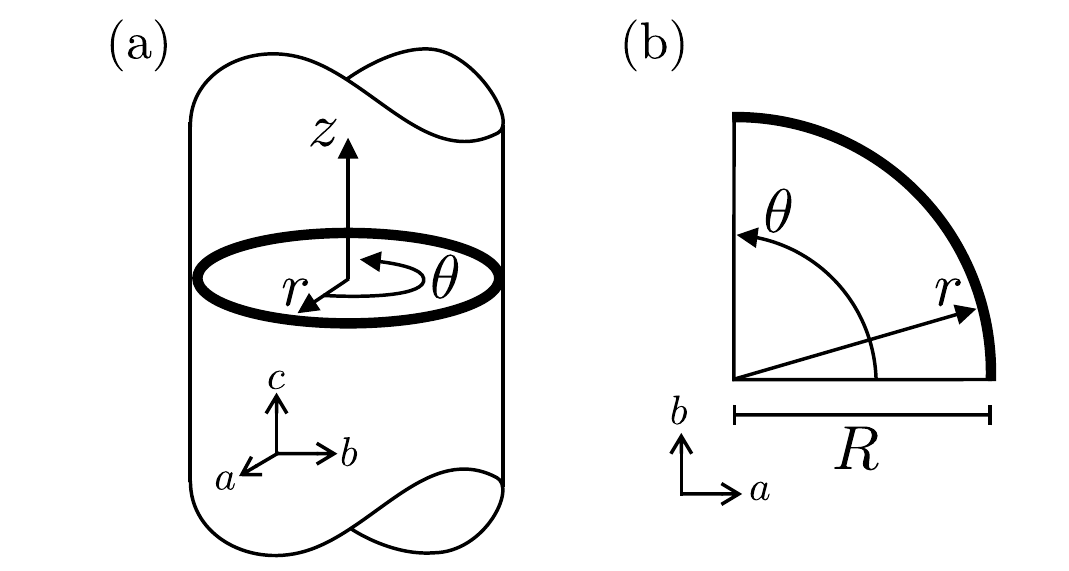}
\caption{\label{fig:geo} Surface (thick line) of the disk geometry, considered cut from an infinite cylinder with translational invariance along the crystal $c$-axis (a). Assuming at least $C_4$ symmetry this is reduced to a quarter disk (b) with $\theta=0$ and $\theta=\pi/2$ corresponding to the crystal $a$- and $b$-axis, respectively.}
\end{figure}

The real-space geometry considered here is a two-dimensional disk cut from an infinite cylinder along the crystal $c$-axis in order to avoid the discussion of boundary effects at the top and bottom faces, as illustrated in Fig.~\hyperref[fig:geo]{\ref*{fig:geo}(a)}. This sample shape includes at once all possible surface orientations with a normal in the basal plane, and additionally allows to study the effect of surface curvature. The radius is chosen such that $R\gg\xi_0,\lambda$. Even in the case of an anisotropic system, at least $C_4 $ symmetry is assumed such that a quarter disk, as shown in Fig.~\hyperref[fig:geo]{\ref*{fig:geo}(b)}, contains all information.

\emph{Polar coordinates:}\quad
Polar coordinates $ (r, \theta )$ are the natural choice for the disk geometry, where $ r \in [0,R] $ and $\theta\in[0,\pi/2)$, defining $\theta=0$ and $\theta=\pi/2$ as the crystal $a$- and $b$-axis, respectively. Assuming translational invariance along the $c$-axis, the corresponding gradient term ($K_5$) in the free energy functional can be ignored such that $\boldsymbol{\eta}(\boldsymbol{r})=\boldsymbol{\eta}(r,\theta)$. The vector potential is $\boldsymbol{A}(\boldsymbol{r})=(A_r(r,\theta),A_\theta(r,\theta),0)= (A_{\perp}(r,\theta),A_{\parallel}(r,\theta),0) $ for a specific choice of gauge and the magnetic field only has a $z$-component $\boldsymbol{B}(\boldsymbol{r})=B_z(r,\theta)\boldsymbol{\hat{z}}$. 

For the order parameter it is also convenient to turn to the polar representation $ \boldsymbol{\eta} =(\eta_r,\eta_\theta) $, which is rewritten using the components being perpendicular and parallel to the surface, $ \eta_r \to \eta_\perp $ and $ \eta_\theta \to \eta_\parallel $, with
\begin{align}\label{eq:para}
\boldsymbol{\eta}&=(\eta_\perp,\eta_\parallel)  = ( \boldsymbol{n} \cdot \boldsymbol{\eta}, \left(\boldsymbol{n} \times \boldsymbol{\eta} \right) \cdot\boldsymbol{\hat{z}}) \\
&= (\eta_x \cos\theta + \eta_y \sin\theta, -\eta_x \sin\theta + \eta_y \cos\theta).\nonumber
\end{align}
Within this notation the real-space phase winding number $N=\pm1$ for the two degenerate chiralities is introduced as $ \boldsymbol{\eta}=(\eta_{\perp},\eta_{\parallel})=\eta_{\rm b} (u,i v)e^{iN\theta}$ with the still complex functions $u$ and $v$ of $ (r, \theta) $. In the homogeneous bulk phase, $ (\eta_\perp,\eta_\parallel)= \eta_{\rm b} (1, \pm i )e^{ \pm i\theta}$. For an isotropic system $u=u(r)$ and $v=v(r)$ have no angular dependence.

\emph{Boundary conditions:}\quad
For the disk geometry with $\boldsymbol{n}=\boldsymbol{\hat{r}}=(\cos\theta,\sin\theta)$ and using the above representation for the order parameter, the surface term from Eq.~(\ref{eq:surf}) is rewritten as
\begin{align}
\mathcal{F}_{\mathrm{surf}}&[\eta_\perp,\eta_\parallel]=\int R\mathrm{d}\theta\bigg[g_1|\boldsymbol{\eta}|^2\\
&+\left(g_3\cos^2(2\theta)+\frac{g_4}{2}\sin^2(2\theta)\right)\left(|\eta_\perp|^2-|\eta_\parallel|^2\right)\nonumber\\
&-\left(g_3-\frac{g_4}{2}\right)\sin(2\theta)\cos(2\theta)\left(\eta_\perp^{*}\eta_\parallel+\eta_\perp\eta_\parallel^{*}\right)\bigg]\nonumber.
\end{align}
In order to reduce the number of variable system parameters, we insert the relation $ g_3 = g_4/2 $, strictly valid only for isotropic symmetry. This leads to the simple form
\begin{align}
\mathcal{F}_{\mathrm{surf}}[\eta_\perp,\eta_\parallel]=\int R\mathrm{d}\theta\Big[(g_1+g_3)|\eta_\perp|^2+(g_1-g_3)|\eta_\parallel|^2\Big] ,
\end{align}
which we will use in the following for the boundary conditions.
In the free energy functional, it is implemented straightforwardly by introducing an $r$-dependent critical temperature
\begin{align} \label{eq:tceff}
T_{ci}^\mathrm{eff}(\boldsymbol{r})=T_c-T_{ci}\delta\left(r-R\right)
\end{align}
for the two order parameter components, with
\begin{align} \label{eq:tc}
T_{c\perp}=\frac{g_1+g_3}{a},\quad T_{c\parallel}=\frac{g_1-g_3}{a},
\end{align}
assuming that for $ r > R $ there is vacuum. For the magnetic field, the boundary condition is $ B_z(R,\theta) = 0 $.

\subsection{Computational Method}

The GL free energy functional is minimized numerically using a one-step relaxed Newton-Jacobi method\cite{gardan:1985,ortega:2000,piette:2004,etter:2017}, for which the indirect boundary conditions through the effective critical temperatures can be implemented straightforwardly by changing the critical temperature at the surface of the disk according to the values of $g_i$ described in the next section. Both the order parameter and the vector potential are discretized on a regular polar grid with uniform step size and with $n=m=100$ mesh points for the radial and azimuthal coordinate, respectively. 
The fact that the tetragonal symmetry is kept in general, including the $ C_4 $ rotation, allows a restriction of our calculation to a quarter disk 
subject to periodic boundary conditions in the angular direction. In the radial direction, the origin is excluded by a circle of radius $ r_\mathrm{min}=10\xi_0$. The disk radius is chosen to be $ R = 40 \xi_0$, which for Sr$_2$RuO$_4$ would be a few $\mu$m in size\cite{maeno:2012}. The gauge is chosen by fixing the order parameter values $\boldsymbol{\eta}=\eta_0(1,+i)$ and the vector potential $\boldsymbol{A}=0$ at $r_\mathrm{min}$, far enough away from the surface such that all quantities are constant. In certain limits analytical arguments allow us a comparison with the numerical results which in all cases agree very well. Also, the results from previous theoretical work are recovered where applicable.

\section{Range of system parameters}\label{sec:scan}

To explore the behavior of the surface magnetic flux pattern under various conditions, two essential parameter sets have been identified which we will scan systemtically. The ratio $ K_1/K_2 $ incorporates effects of the lattice on the electronic structure, while $ g_1 $ and $ g_3 $ cover a range of different surface types. Based on these parameters the different types of systems are defined and compared with previous studies.

Motivated by the results in Ref.~[\onlinecite{bouhon:2014}], the ratio $ K_1/K_2 $ is scanned between the isotropic limit $K_1/K_2=3$ through $K_1=K_2$ to the `inverted' ratio $K_1/K_2=1/3$. In this way the feature of current reversal\cite{bouhon:2014} can be reproduced for certain boundary conditions.
While changing $ K_1/K_2 $ has been discussed before, the range of  $K_1<K_2$ has not been studied within the GL approach so far\cite{ashby:2009,lederer:2014}. For our analysis, the difference $(K_1-K_2)/K$ is varied rather than the ratio, as it turns out that the magnetic flux pattern depends on this difference, which can be seen for example from Eq.~(\ref{eq:jtspecp4}).

\begin{figure}[t]
\includegraphics[width=0.9\columnwidth]{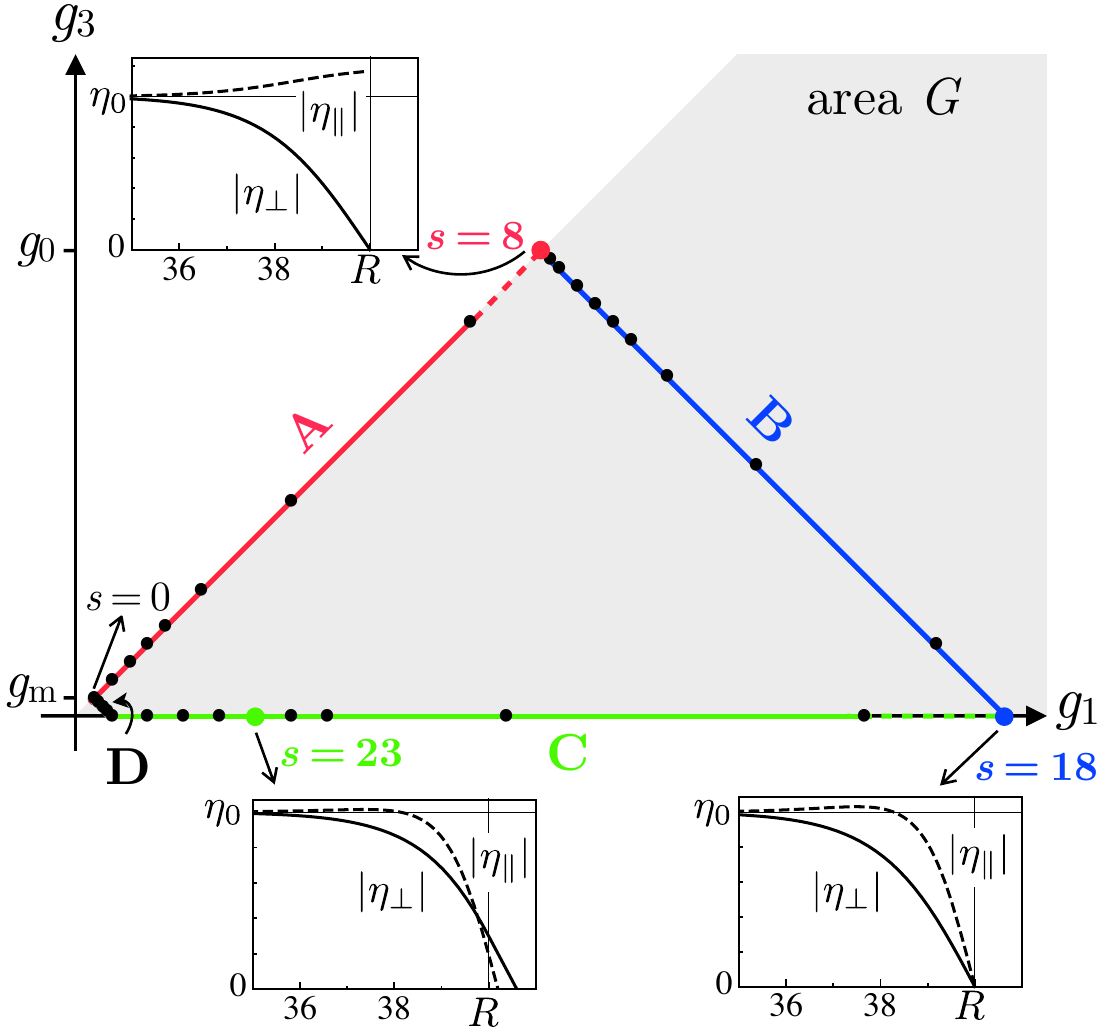}
\caption{\label{fig:surftype} The area $G$ of the surface coefficients $g_1$ and $g_3$ covering the effective critical temperatures $T_{c\perp}^\mathrm{eff} \leq T_{c\parallel}^\mathrm{eff}\leq T_c$. The numerical analysis includes 31 cases $s$ (black dots) on the three ranges A, B, and C, and on D to exclude the origin. Three representative surface types are highlighted.}
\end{figure}

Implementing the boundary conditions through the effective critical temperatures at the surface given in Eq.~(\ref{eq:tc}), the condition $ T_{ci}\geq 0 $ describes the destructive effect of the surface on superconductivity in general and leads to $ g_1 \geq | g_3| $. The restriction $T_{c\perp} \geq T_{c\parallel}$ further excludes scenarios where the parallel component of the order parameter would be suppressed more and leads to $ g_3 \geq 0 $. The values $ g_1 = g_3 = 0 $ describe a `virtual boundary' in the interior of the superconductor without any effect on the order parameter. These requirements result in the area $G=\left\{g_i\ |\ (g_1\geq g_3\geq 0)\ \wedge\ (g_1+ g_3>0)\right\}$, corresponding to the region shaded gray in Fig.~\ref{fig:surftype}.

In the following, we focus on surface types along the ranges A, B, and C, the bounds of the area $G$, and introduce an additional range D to circumvent the origin, all indicated in Fig.~\ref{fig:surftype}. For the numerical analysis, the results are systematically computed for 31 cases $s\in\{0,\dots,30\}$, indicated by black dots, and whose actual values are listed in the supplemental material\cite{suppmat}. Three representative cases of particular interest are highlighted and illustrated by insets. The range A along the identity $ g_1 = g_3 $ corresponds to surfaces with specular scattering. With $T_{c\parallel}=0$, the order parameter component $ \eta_{\parallel} $ is little affected, as in Ref.~[\onlinecite{matsumoto:1999}], while $T_{c\perp}$ grows with increasing $g_i$ and leads to a progressive suppression of $ \eta_{\perp} $ at the surface, for which $ \eta_{\parallel} $ is even slightly enhanced. The case $ s=8 $ corresponds to the limit $ g_{1,3} = g_0 \to \infty $ where the extrapolation length $ l_{\perp} \ll \xi_0 $ such that $ \eta_{\perp} (R,\theta) \rightarrow 0 $. For the numerical treatment, whenever $T_{ci}^\mathrm{eff}\rightarrow-\infty$, the direct boundary conditions $\eta_i(R)=0$ are implemented instead, as described in the supplemental material\cite{suppmat}.
The range B, defined by $ g_1+g_2= 2 g_0 $, keeps $  \eta_{\perp} $ fully suppressed at the boundary, whereby also $ \eta_{\parallel}$ is reduced with decreasing $ g_3 $. For the case $ s=18 $, where $ g_3 = 0 $, both order parameter components vanish completely at the boundary. In range C, both order parameter components remain reduced but with finite extrapolations lengths. Both B and C describe effects due to surface roughness and diffuse scattering.

Before systematically analyzing the solution to our GL model, we briefly comment on the previous theoretical discussion of surface types. The commonly investigated situations are planar surfaces with specular scattering. Assuming full rotation symmetry around the $c$-axis, i.e. an isotropic system, a peak value for edge current induced fields of the order of 1 mT has been obtained\cite{matsumoto:1999}, which is often used in experimental investigations in Sr$_2$RuO$_4$ as a reference\cite{kirtley:2007,hicks:2010,curran:2014}. This case corresponds to our results at $s=8$ for $ K_1/K_2 = 3$ with $ \eta_{\perp} = 0 $ at the boundary. While our approach also covers ranges with finite $ \eta_{\perp} $, most treatments impose the condition of its full suppression  following the pioneering work by Ambegaokar, de Gennes and Rainer \cite{ambegaokar:1974}. These studies are therefore located along the line B\cite{huang:2012,nagato:1998,ashby:2009}. Ashby and Kallin\cite{ashby:2009} considered the special case of `diffuse' scattering with an extrapolation length $ l_{\perp} = 0.54 $ following the results of Ref.~[\onlinecite{ambegaokar:1974}], which is  $ s\approx11 $ in our discussion. Their analysis also includes `full pair breaking' at the surface\cite{ashby:2009}, which here is $s=18$. While these works have considered planar surfaces, disk geometries with rather small radius have been studied by Huang and Yip\cite{huang:2012} and Suzuki and Asano\cite{suzuki:2016}. On the other hand, normal metallic surface layers lead to a further type of boundary condition with finite extrapolation lengths for both order parameter components, as discussed previously by Lederer \emph{et al.}\cite{lederer:2014}. These authors considered also the special case of $ K_1 = K_2 $ and identical extrapolation lengths, $ l_{\perp} = l_{\parallel} $.  Very recently, Bakurskiy \emph{et al.}\cite{bakurskiy:2017} have extended the discussion to different surface layers of varying roughness and metallic behavior. These two works are covered by our analysis of range C.

\section{Detailed Analysis}\label{sec:res}

In this section, the spontaneous magnetic flux distribution generated by the surface currents is systematically analyzed. Starting with the isotropic limit, the ratio $ K_1/K_3 = 3 $ is fixed while the boundary conditions are changed. Then, restricted to specular scattering at the surface, anisotropy is introduced by varying $ K_1/K_3 $ between $ 3 $ and $ 1/3 $. Eventually, all surface types $s$ and all ratios $K_1/K_2$ are scanned and discussed systematically. The full set of numerical results can be found in the supplemental material\cite{suppmat}, while a selective analysis comprising the most essential features is given in this section.

\subsection{Different surface types in the isotropic limit}\label{sec:res:iso}

\begin{figure}[t]
\center
\includegraphics[width=0.9\columnwidth]{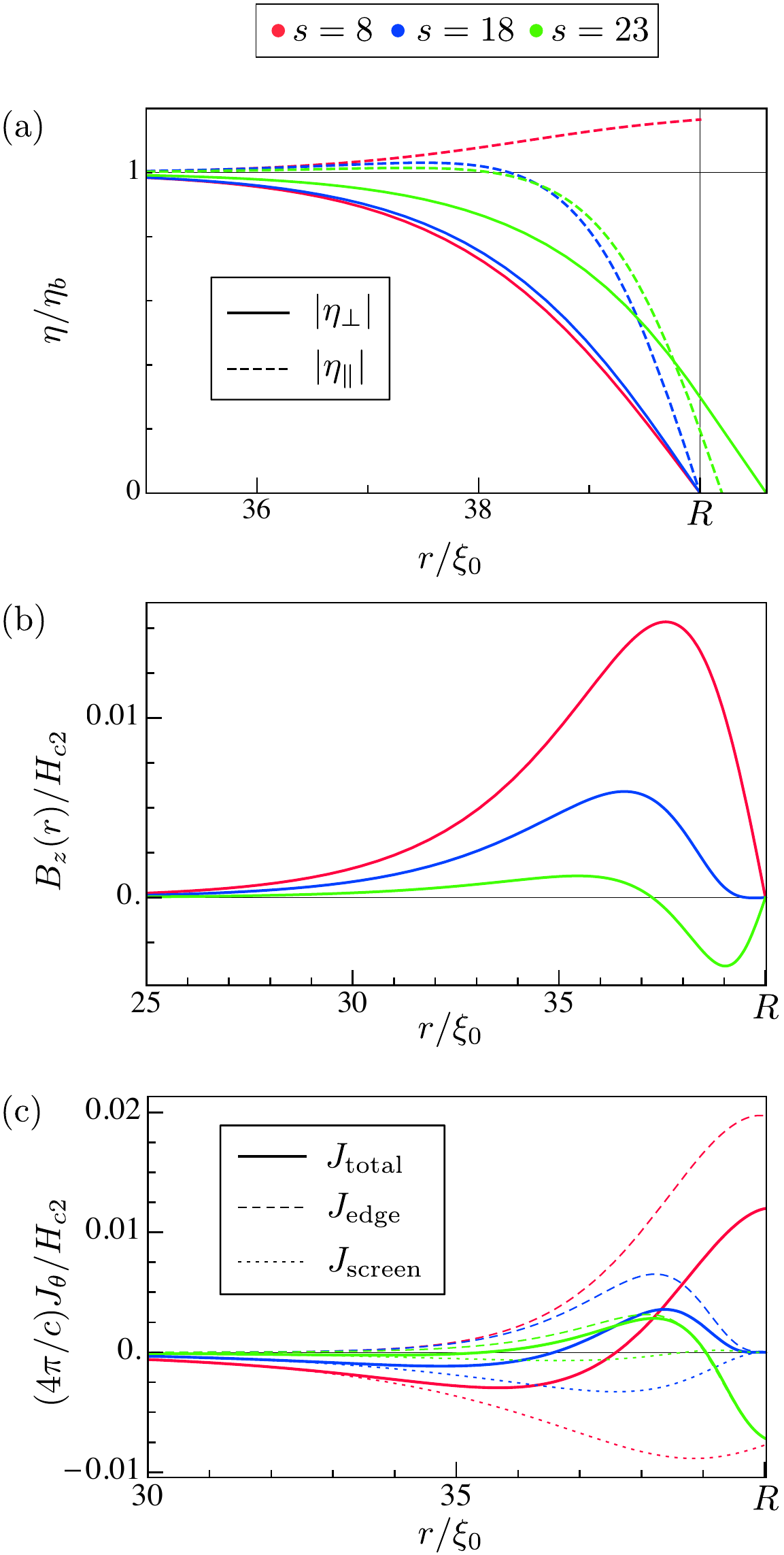}
\caption{\label{fig:iso} Numerical results for the three surface types specular scattering ($s=8$), full pair breaking ($s=18$) and a rough surface ($s=23$) in the isotropic limit. (a) The absolute value of the order parameter components $\eta_{\perp}$ and $\eta_{\parallel}$. (b) The magnetic flux $B_z$. (c) The total current at the surface $J_{\parallel}$, separated into  the `edge' and `screening' part.}
\end{figure}

In the isotropic limit, $K_1/K_2=3$, all quantities such as the order parameter and the magnetic flux pattern have no angular dependence. The numerical results are shown in Fig.~\ref{fig:iso} for the three representative surface types highlighted in Fig.~\ref{fig:surftype}: specular scattering ($s=8$), full pair breaking ($s=18$) and the case of a rough surface with moderate pair breaking ($s=23$). 

The order parameter components $\eta_{\perp}(r) $ and $\eta_{\parallel}(r) $ are displayed in Fig.~\hyperref[fig:iso]{\ref*{fig:iso}(a)}. In the absence of angular dependence the global phase can be set constant with $ \Im \eta_{\perp} = 0 $ and $ \Re{\eta_{\parallel}}=0 $.
It is worth noting that for specular scattering $\eta_{\parallel}$ has an increasing slope at $r=R$, unlike on a planar surface, because the curvature supports this order parameter component, as demonstrated in Eq.~(\ref{eq:app:spec:slopets}) in Appendix~\ref{app:bc}, also in agreement with Suzuki and Asano\cite{suzuki:2016}. For $s=18$ both components are suppressed to zero and for $s=23$ they are reduced at the surface with a finite extrapolation length on the order of the coherence length.

The magnetic field $B_z(r)$ due to edge currents is shown in Fig.~\hyperref[fig:iso]{\ref*{fig:iso}(b)}. Specular scattering leads to a single positive peak near the surface, as found in Refs.~[\onlinecite{matsumoto:1999}] and [\onlinecite{ashby:2009}]. A similar peak structure appears also for $ s=18 $, however, with a reduced magnitude, in agreement with Ref.~[\onlinecite{ashby:2009}]. A qualitative change occurs for $ s=23$ where $B_z$ changes sign, starting negative near the surface and crossing to positive at a distance on the order of the coherence length. Interestingly, the total flux of this case is essentially zero. Note that this corresponds to a node of the $s$-dependence of the magnetic flux, as $ s=22 $ and $s=24 $ have a finite total flux of opposite sign (see Fig.~\hyperref[fig:pd]{\ref*{fig:pd}(b)}). 

The surface current $J_{\parallel} $ is depicted in Fig.~\hyperref[fig:iso]{\ref*{fig:iso}(c)}, while the radial current $J_{\perp} $ vanishes everywhere in the isotropic limit. It is illustrative to separate the current into two parts, the driving current due to the edge, $ J_{\rm edge} $, and the screening current to suppress the magnetic field inside the disk, $ J_{\rm screen} $, for the definition see Eq.~(\ref{app:eq:jparts}) in Appendix~\ref{app:bc}. The spatial extension of these currents has different length scales, the coherence length for  $ J_{\rm edge} $ and the London penetration depth for $ J_{\rm screen} $. Note that these lengths are similar due to the small Ginzburg-Landau parameter $ \kappa = 2.6 $. Again, the driving currents for $s=8$ and 18 are both positive, but the latter is reduced compared to the former and vanishes at the surface, as shown by Eq.~(\ref{app:sub:fpb}) in Appendix~\ref{app:bc}. The current for $ s=23 $ has a sign change and correspondingly the screening current is strongly reduced.

\begin{figure}[t]
\includegraphics[width=0.9\columnwidth]{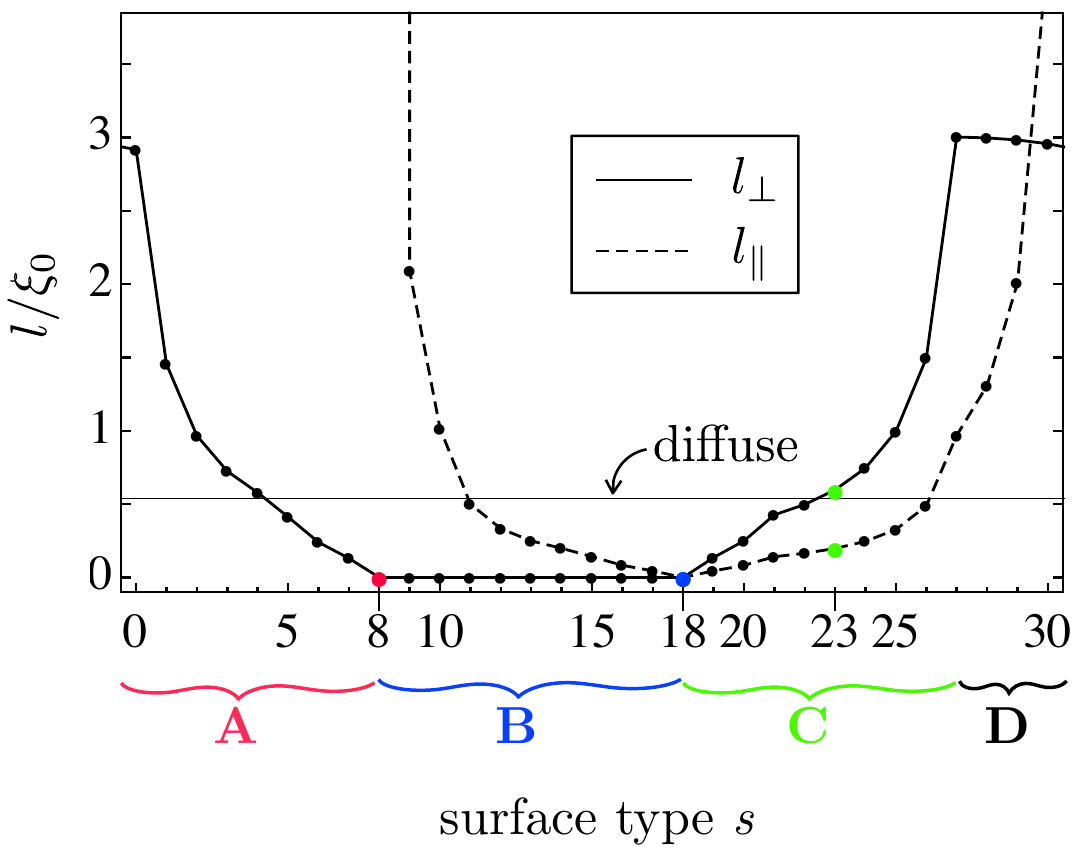}
\caption{\label{fig:extrapol} Extrapolation lengths $l_\perp$ (solid) and $l_\parallel$ (dashed) of the order parameter components in the isotropic limit for all surface types. The ranges A, B, C, and D, the three representative cases $s\in\{8,18,23\}$ (see Fig.~\ref{fig:surftype}), as well as the specific extrapolation length $l_\mathrm{diff}=0.54$ are indicated.}
\end{figure}

In the isotropic limit it is also interesting to consider 
the extrapolation lengths $ l_{\perp} $ and $ l_{\parallel} $ for $ \eta_{\perp} $ and $ \eta_{\parallel} $, respectively, which can be extracted from the numerical results and shown in Fig.~\ref{fig:extrapol}.  For the specular scattering surface types in range A (see Fig.~\ref{fig:surftype}), $l_{\parallel}$ diverges, while $l_{\perp}$ decreases as $\eta_{\perp}$ is progressively suppressed at the surface starting at $s=0$ until $l_{\perp}=0$ for $s=8$. Within range B, $ \eta_{\perp} $ remains zero at the surface such that $ l_{\perp} = 0 $. The gradual suppression of $ \eta_{\parallel} $ leads to a decrease of $ l_{\parallel} $ until it vanishes at $ s=18 $ with full pair breaking. For range C, both extrapolation lengths increase again continuously with $ 3 l_{\parallel} \sim l_{\perp} $ due to the difference in coherence length for the two order parameter components, see Eq.~(\ref{eq:app:iso}) in Appendix~\ref{app:bc}. Along range D, $ l_{\perp} $ basically remains constant, while $ l_{\parallel} $ grows to eventually diverge again at $ s=0$. 

The specific extrapolation length $l_{\mathrm{diff}}=0.54$ is also indicated, which describes the `diffuse' scattering as derived by Ambegaokar, de Gennes and Rainer\cite{ambegaokar:1974} and treated within a GL approach by Ashby and Kallin\cite{ashby:2009}. Our boundary condition $s=11$ is very close to this case and the corresponding results are in agreement with Ref.~[\onlinecite{ashby:2009}].

\subsection{Specular scattering and anisotropy}
\label{sec:spec}

\begin{figure}[t]
\includegraphics[width=0.9\columnwidth]{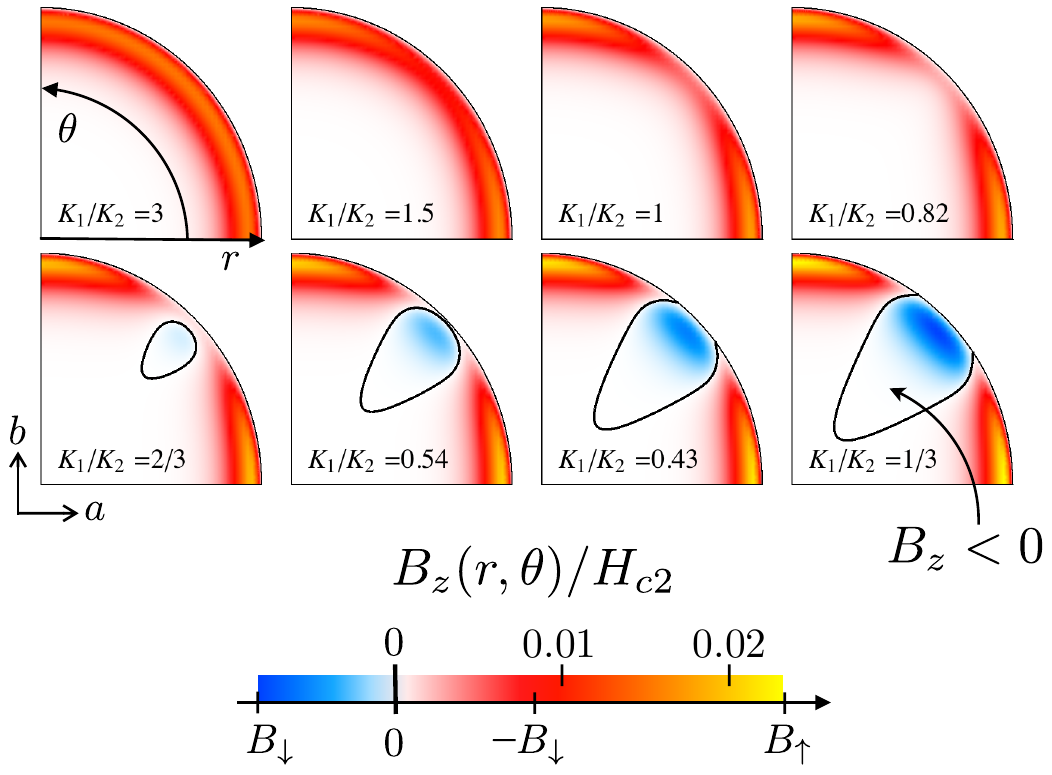}
\caption{\label{fig:spec_all} The magnetic flux $B_z$ for the specular scattering surface type $s=8$ at different ratios $K_1/K_2$. Negative fluxes are blue, with deep blue being the minimum $B_\downarrow$, and encircled by a thick black line. The corresponding positive value $-B_\downarrow$ is colored deep red and the maximum $B_\uparrow$ is yellow. The subplot $K_1/K_2=1/3$ (bottom right) is enlarged in Fig.~\ref{fig:spec_inv_vec}.}
\end{figure}

Turning $K_1/K_2$ away from 3 introduces anisotropy such that all quantities acquire angular dependence. 
Here we restrict our discussion to the specular scattering surface type $s=8$.
The magnetic flux pattern for different ratios $K_1/K_2$ is displayed in Fig.~\ref{fig:spec_all}, starting with the isotropic limit $K_1/K_2=3$ (upper left) to $K_1/K_2=1/3$ (lower right). Due to the assumed $C_4$ symmetry, it is sufficient to display a quarter disk with $ 0 \leq \theta \leq \pi/2 $, where $ \theta = 0 , \pi/2 $ correspond to the crystalline axes of the basal plane. 

\begin{figure}[t]
\includegraphics[width=1.0\columnwidth]{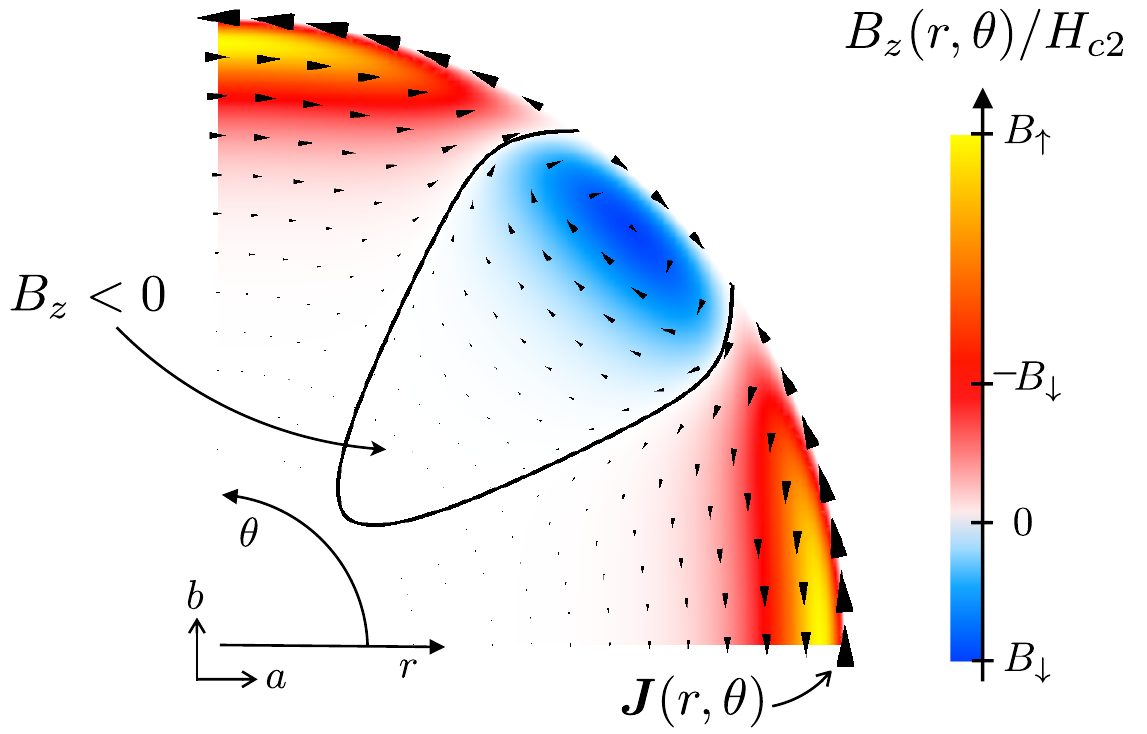}
\caption{\label{fig:spec_inv_vec} Density plot of the magnetic flux $B_z$ and a scaled vector plot of the current $\boldsymbol J$ for the inverted ratio $K_1/K_2=1/3$ and the specular scattering surface $s=8$. The current pointers are scaled by $|\boldsymbol{J}|^{1/3}$ to enhance the small values.}
\end{figure}

The magnetic flux pattern for $K_1/K_2=3$ agrees with the result shown in Fig.~\hyperref[fig:iso]{\ref*{fig:iso}(b)} for $s=8$. Lowering $K_1/K_2$ the flux pattern progressively develops anisotropy. In agreement with Ashby and Kallin\cite{ashby:2009}, the magnetic flux increases along the [10]-directions $\theta= 0 $ and $ \pi/2$. On the other hand, it decreases along the [11]-direction $ \theta = \pi/4 $. When the ratio is $K_1/K_2\approx2/3$ in this direction, a slightly negative flux region appears (encircled by a black line) which expands upon decreasing $ K_1/K_2 $. This behavior is consistent with the observation of current reversal by Bouhon and Sigrist within a lattice BdG approach for a planar surface along the [11]-direction\cite{bouhon:2014}. The current pattern is shown in the more detailed plot for the `inverted' ratio $K_1/K_2 = 1/3$ in Fig.~\ref{fig:spec_inv_vec}, where the surface currents run in a clockwise direction for $\theta= 0 $ and $ \pi/2$, and anti-clockwise for $ \theta = \pi/4 $. Interestingly, the critical ratio for the onset of the current reversal is lower for the curved than for a planar surface, as discussed for Eq.~(\ref{app:eq:jspec}) in Appendix~\ref{app:bc}. Increasing the disk radius to $R=240\xi_0$, we observe the negative flux and surface currents at higher ratios, closer to $K_1=K_2$.

\subsection{Full scan of parameters}\label{sec:fullscan}

\begin{figure}[t]
\includegraphics[width=0.95\columnwidth]{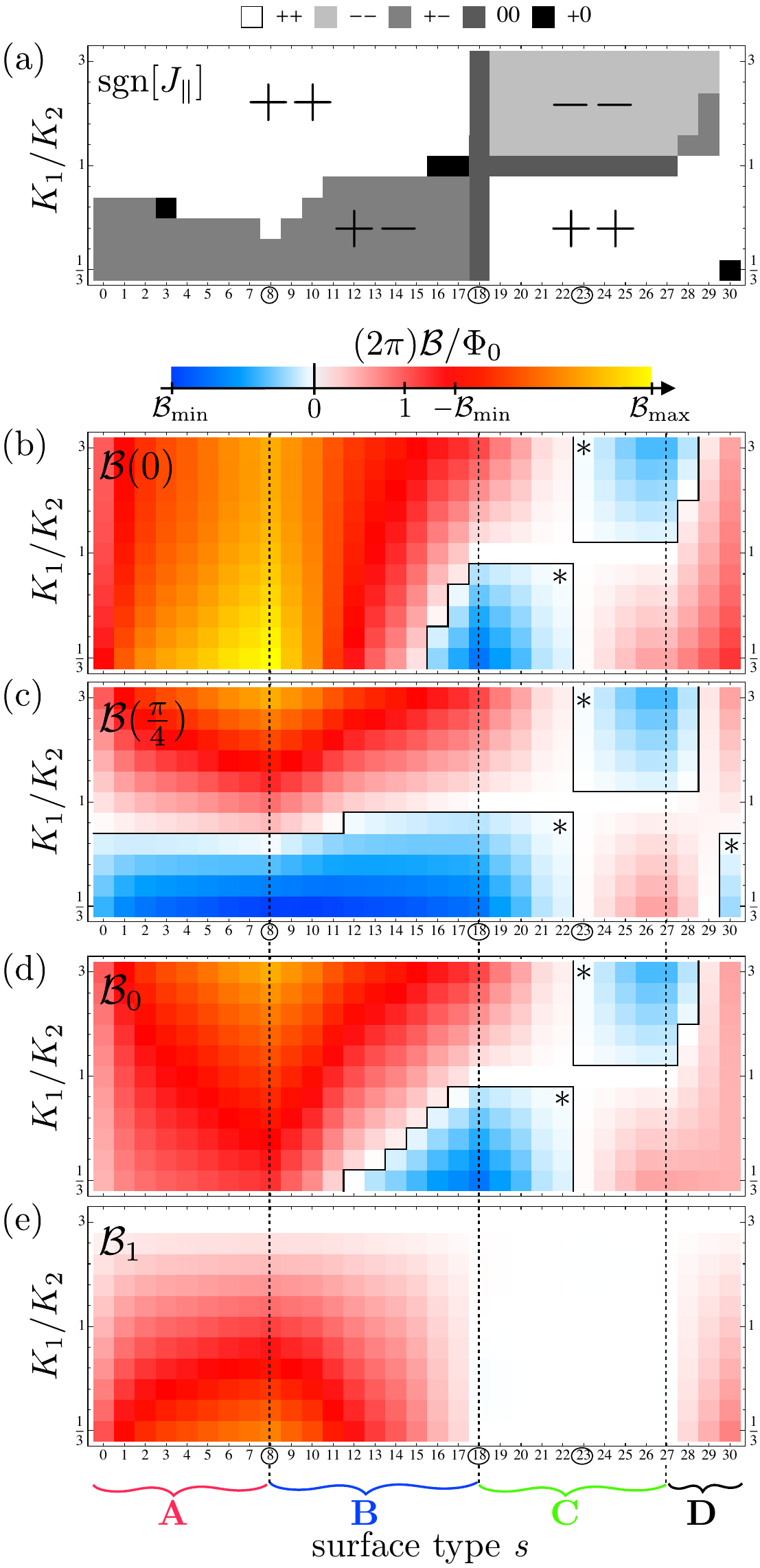}
\caption{\label{fig:pd} 
Extracted quantities: (a) the current directions $\{\mathrm{sgn} J_{\parallel}(0),\mathrm{sgn}J_{\parallel}(\pi/4)\}$; the radially integrated flux $\mathcal{B}(\theta)$ at (b) $\theta=0$ and (c) $\theta=\pi/4$; (d) the total flux $\mathcal B_0=\Phi/(2\pi)$, and (e) the first component of the Fourier analysis $\mathcal B_1$. The color scale ranges from the overall minimum $ \mathcal B_\mathrm{min}<0 $ (deep blue) through the corresponding positive value (deep red) to the overall maximum $ \mathcal B_\mathrm{max} >0$ (yellow). Areas of negative $\mathcal B$ are encircled by a black line and labeled $*$.}
\end{figure}

Finally, the flux pattern is analyzed for the whole range of surface types and ratios $K_1/K_2$. Selected quantities extracted from the numerical results are presented in Fig.~\ref{fig:pd} as density plots with $s$ along the horizontal and $ K_1/ K_2 $ along the vertical axis. Note that the vertical scale is non-linear because equal steps are taken in $(K_1-K_2)/K $. A useful symmetry relation for the magnetic field is explained in Appendix~\ref{app:symm},
\begin{align}
\label{eq:oddB}
B_z[K_1/K_2,g_1,g_3=0]=-B_z[K_2/K_1,g_1,g_3=0].
\end{align}
In particular, it establishes the observation that  the magnetic flux vanishes for $ K_1 = K_2 $ and $ g_3 = 0 $ (range C), see also Ref.~[\onlinecite{lederer:2014}].

First, the direction of the surface current for the surface orientations $\theta=0$ and $\pi/4$ is analyzed in Fig.~\hyperref[fig:pd]{\ref*{fig:pd}(a)}, considering only the signs $\{\mathrm{sgn}[ J_{\parallel}(\theta=0)],\mathrm{sgn}[J_{\parallel}(\theta=\pi/4)]\}$.
For surface types A and B, the current $ J_{\parallel}(\theta =0) $ is positive for all ratios $K_1/K_2$. On the other hand,  the current $J_{\parallel} (\theta = \pi/4) $ changes sign at a value $ K_1/K_2 < 1 $ varying moderately with the surface type $s$. At $ s=18 $, the surface current vanishes for all ratios $K_1/K_2 $, see Eq.~(\ref{app:sub:fpb}) in Appendix~\ref{app:bc}, with the onset pushed inside the disk as illustrated in Fig.~\hyperref[fig:pd]{\ref*{fig:pd}(c)}. For range C, the surface current starts in the isotropic limit with negative values for both surface orientations. In agreement with the symmetry relation in Eq.~(\ref{eq:oddB}), the current vanishes identically for the whole disk at $ K_1=K_2$ and changes direction for $K_1<K_2$.

More insight into the behavior of the magnetic flux pattern is obtained by considering the angular dependence after integrating over the radial direction,
\begin{equation}
\mathcal B(\theta)=\int r\mathrm{d}rB_z(r,\theta) .
\end{equation}
This is analyzed successively in Figs.~\hyperref[fig:pd]{\ref*{fig:pd}(b)-(e)}, where the same color scale is used in all subplots for an easy comparison, with $\mathcal{B}_\mathrm{min}$ and $\mathcal{B}_\mathrm{max}$ being the overall extremal values, occurring as discussed below.
First, we discuss $ \mathcal B(\theta) $ for the orientations $ \theta = 0 $ and $ \pi/4 $ in Figs.~\hyperref[fig:pd]{\ref*{fig:pd}(b)} and \hyperref[fig:pd]{(c)}, respectively. In range A and B, $ \mathcal B(0) $ is positive with the overall maximum $ \mathcal B_\mathrm{max} $ reached for the specular scattering case $ s=8 $ and the most pronounced anisotropy $ K_1/K_2 = 1/3 $. The characteristic feature of $ \mathcal B(\pi/4) $ is the sign change at a ratio $ K_1 / K_2 < 1 $. The overall minimum $ \mathcal B_\mathrm{min} $ ($ < 0$) is reached for $s=8$ and $K_1/K_2=1/3$. The same pattern appears for both  $ \mathcal B(0) $ and  $ \mathcal B(\pi/4) $ in range C, with a vanishing flux when $ K_1 = K_2 $, based on the symmetry relation in Eq.~(\ref{eq:oddB}). For $s=23$, the flux vanishes for all ratios $K_1/K_2$.

Next, we examine the angular dependence of $\mathcal B(\theta)$ using the Fourier decomposition
\begin{equation}
\mathcal B(\theta) = \sum_{n=0}^{\infty} \mathcal B_n \cos (4n \theta).
\end{equation}
The total magnetic flux through the disk is $ \Phi = 2 \pi  \mathcal B_0$ and shown in Fig.~\hyperref[fig:pd]{\ref*{fig:pd}(d)}. The lowest order angular dependence $\mathcal{B}_1$ is shown in Fig.~\hyperref[fig:pd]{\ref*{fig:pd}(e)}. Typically, we find $ |\mathcal B_{n\geq 2}| \ll  \{|\mathcal B_0|, | \mathcal B_1| \}$, such that higher orders can be neglected here.
For $s$ in A and B, naturally $  \mathcal B_0 $ dominates in the more isotropic ($K_1>K_2$) and $  \mathcal B_1 $ in the more anisotropic ($K_1<K_2$) regime. Interestingly, the total flux $  \mathcal B_0$ can also become negative in range B for a sufficiently pronounced anisotropy. The strongest angular dependence is obtained again for specular scattering at $s=8 $ and the maximal anisotropy $K_1/K_2=1/3$. The magnetic flux pattern in range C is essentially isotropic $  \mathcal B_1 \approx 0 $. Therefore, independent of the surface orientation, the magnetic flux vanishes for the case $ s=23 $ for all $ K_1/K_2 $ and for $ K_1 = K_2 $ for all $s \in \{18, \dots, 27\} $, the latter again based on the symmetry relation Eq.~(\ref{eq:oddB}).

We conclude that the case of specular scattering gives rise to the largest magnetic fields. This `upper bound' corresponds to the situation discussed in Ref.~[\onlinecite{matsumoto:1999}] which is often taken as a reference for the magnitude of the spontaneous magnetic fields at the surface. It can be enhanced somewhat by anisotropy. We also reproduce the current reversal obtained by a lattice BdG approach\cite{bouhon:2014}. Most surprising is the fact that the magnetic fields can be suppressed for all surface orientations in a wider range when the boundary conditions are sufficiently destructive.
In particular, the vanishing total flux and the current reversal are not isolated and fine-tuned results, but appear for a larger range of parameters.

\section{Impurity}\label{sec:imp}

Besides boundaries, impurities also distort the order parameter in unconventional superconductors. Analogous to the surface, for chiral $p$-wave superconductors this leads to spontaneous currents circling around the impurity, inducing a local magnetic field. Within our GL framework the presence of a single isolated bulk impurity at $ \boldsymbol{r}_\mathrm{imp} $ is easily implemented by a correction to the critical temperature,
\begin{equation}
T_{c}^\mathrm{eff}=T_c\big(1-\tau_\mathrm{imp} \delta(\boldsymbol{r}-\boldsymbol{r}_\mathrm{imp})\big),
\end{equation}
where $\tau_\mathrm{imp}>0$ corresponds to the depletion strength of the order parameter, see also Ref.~[\onlinecite{okuno:1999}]. As a quasi two-dimensional superconductor is considered, the coherence length along the $z$-direction is very short, such that the driving currents are well concentrated in the layer where the impurity is located and screening is mainly due to currents parallel to the basal plane. For simplicity, the discussion is restricted to a two-dimensional system. For the numerical analysis, a square with side length $ L = 30 \xi_0 $ is considered, with the point-like impurity located in the center. In analogy to the discussion above, the ratio $K_1/K_2$ again introduces anisotropy, while now the strength $\tau_\mathrm{imp}$ is the additional scanning parameter.

\begin{figure}[t]
\includegraphics[width=1.0\columnwidth]{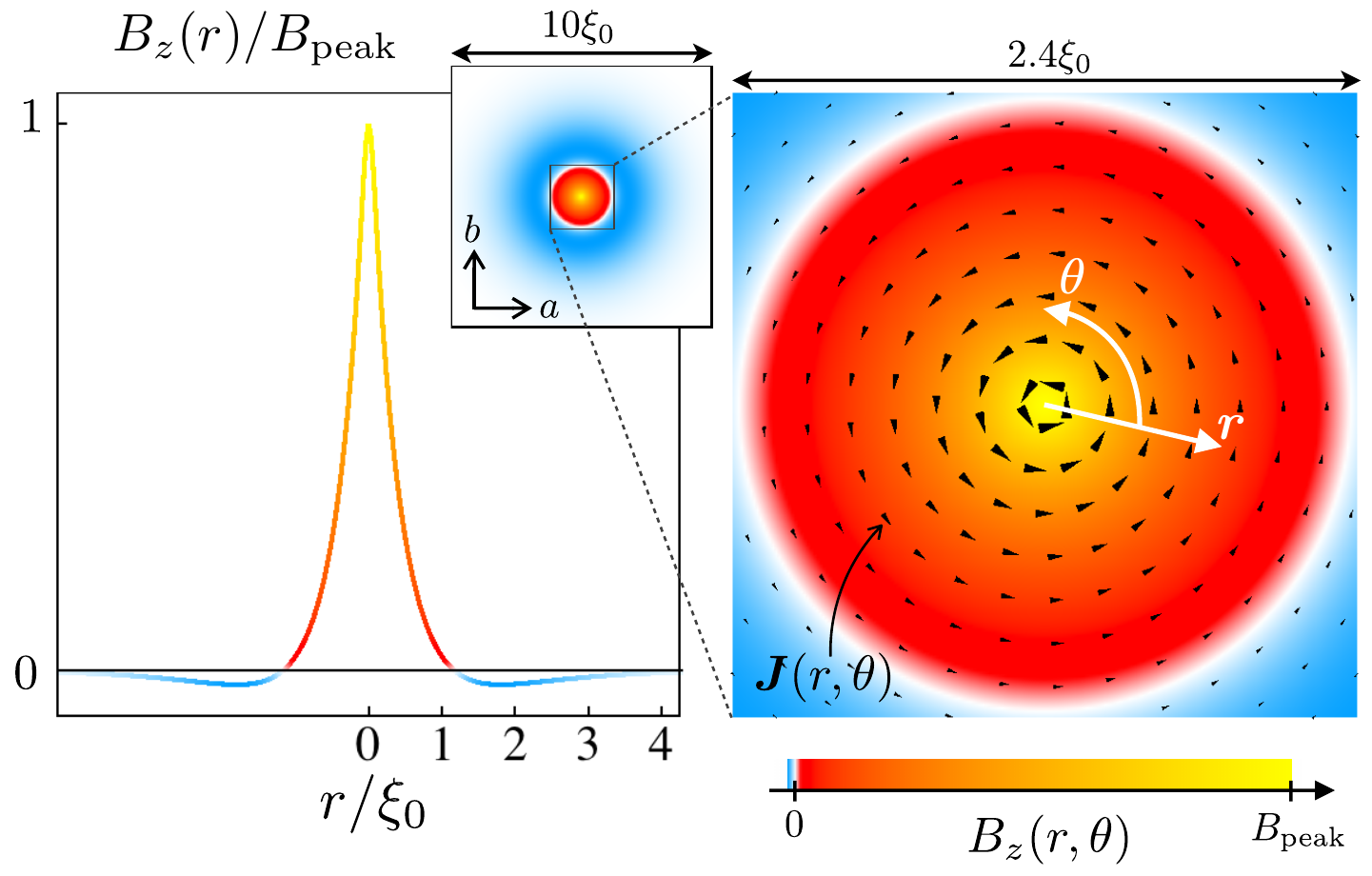}
\caption{\label{fig:1imp} Magnetic flux pattern $B_z$ around an isolated impurity with strength $\tau_\mathrm{imp}=50\xi_0^2$ where the order parameter is fully suppressed, and in the isotropic limit. The colors red to yellow indicate increasing positive flux, while blue is the much smaller negative screening flux at larger distances. Again, the pointers are scaled by $|\boldsymbol{J}|^{1/3}$ to enhance small values.}
\end{figure}

\begin{figure}[t]
\includegraphics[width=1.0\columnwidth]{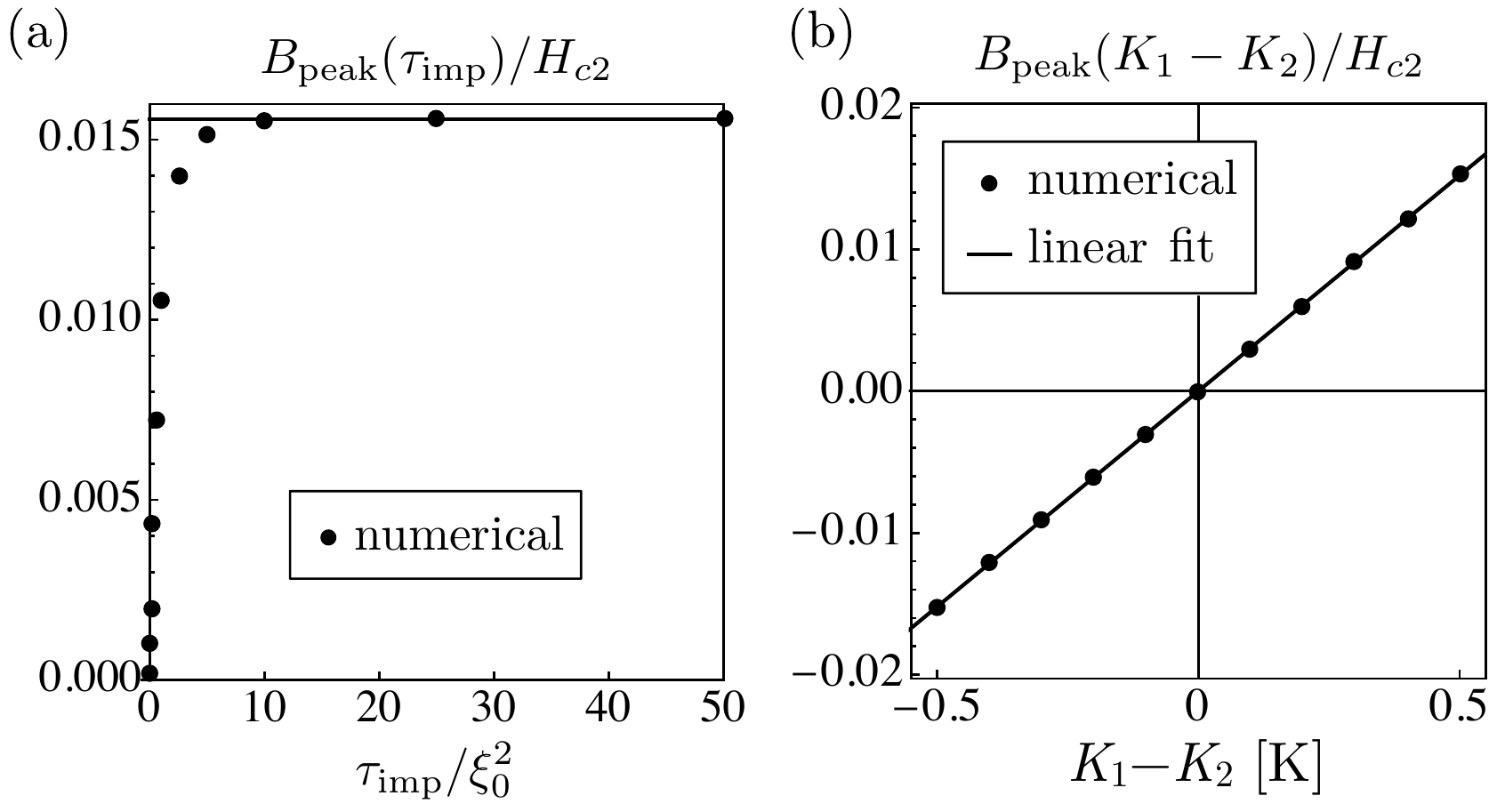}
\caption{\label{fig:imp_k} Dependence of the magnitude of the flux peak $B_\mathrm{peak}$ on (a) the strength $\tau_\mathrm{imp}$ in the isotropic limit; and on (b) the ratio $K_1/K_2$ ($K_1-K_2=0.5 K$ corresponds to the isotropic limit) for an impurity with a strength $\tau_\mathrm{imp}=10\xi_0^2$ where the order parameter is fully suppressed and $B_\mathrm{peak}$ has saturated.}
\end{figure}

The magnetic field and current distribution around the impurity are shown in Fig.~\ref{fig:1imp} for the isotropic limit $K_1/K_2=3$ and for a strength $\tau_\mathrm{imp}$ at which the order parameter is fully suppressed in the center. The magnetic field is strongly peaked at the impurity site $r = 0$ and is compensated by a negative tail at larger distances. The total magnetic flux $\Phi_\mathrm{imp}$ vanishes as required. The magnitude of the flux peak $B_\mathrm{peak}$ depends both on the ratio $K_1/K_2$ and on the strength $\tau_\mathrm{imp}$, as shown in Fig.~\ref{fig:imp_k}. The two effects are considered separately. First, the effect of the strength $\tau_\mathrm{imp}$ is shown in Fig.~\hyperref[fig:imp_k]{\ref*{fig:imp_k}(a)} for the isotropic limit $K_1/K_2=3$. At a large enough value of $\tau_\mathrm{imp}$, both order parameter components are fully suppressed at the site of the impurity, and the peak strength $B_\mathrm{peak}$ saturates.
Second, the effect of the anisotropy from scanning the ratio $K_1/K_2$ is shown in Fig.~\hyperref[fig:imp_k]{\ref*{fig:imp_k}(b)} for a value of $\tau_\mathrm{imp}$ where $B_\mathrm{peak}$ has saturated in the isotropic limit. Equal steps in $(K_1-K_2)/K$ are taken. The odd dependence of $B_\mathrm{peak}$ on $(K_1-K_2)/K$ is in agreement with the symmetry relation from Eq.~(\ref{eq:oddB}), which also applies here. The peak magnitude is essentially linear in $(K_1-K_2)/K$. 

\section{Conclusion}\label{sec:disc}

The aim of our study was to illustrate the variability of the spontaneous magnetic flux pattern $ B_z(r,\theta) $ at the surface of a chiral $p$-wave superconductor. The disk geometry provides a
sample including all surface orientations. The Ginzburg Landau approach allows us to examine a wide range of conditions, in particular, different boundary conditions and the anisotropy of the electronic structure due to the crystal lattice. For the purpose of a systematic scan of these conditions for a given chirality, the relevant `anisotropy' parameter $ K_1 / K_2 $ was introduced and an extensive set of surface types labeled by $s$ was chosen, while ignoring the variability of other parameters in the GL functional. 
We compared our results with those of previous studies and could reproduce these results as special cases of our more extensive discussion. In particular, we verified also the somewhat counter-intuitive current reversal around the direction [110] found in Ref.~[\onlinecite{bouhon:2014}].

Our main result is that the flux pattern is not a universal feature of the topologically non-trivial chiral superconducting phase, but it is extremely sensitive to both the ratio $K_1/K_2$ and the surface type $s$, in combination with the surface orientation. This is apparent immediately on the chart of all flux patterns provided in the supplemental material\cite{suppmat}. The naive picture of a chiral superconductor to behave like an orbital ferromagnet where edge currents are considered as uncompensated circular currents does definitely not 
conform with our findings. Not only the magnitude of $ B_z(r,\theta) $ varies, but it can also be highly anisotropic and even change sign. Only a specular scattering surface in the isotropic limit conforms well with the naive expectation. There is a certain range of conditions where the magnetic flux vanishes completely. Interestingly, the magnetic flux at an impurity follows a behavior similar to the flux at the surface along range C concerning the dependence on $ K_1 / K_2 $. Additionally, however, the peak strength at the impurity is governed by the phenomenological parameter $\tau_\mathrm{imp}$, and saturates when both order parameter components are fully suppressed. 

We also considered the radially integrated flux and found that there are parameter ranges where this quantity vanishes due to cancellation, although the local magnetic fields are non-vanishing. Since the extension of the flux pattern from the surface is limited to a length of order of the London penetration depth, this fact may also be relevant for experimental detection by devices which have a considerably coarser spatial resolution than this. On the other hand, experimental accuracy has improved considerably over the last few years making the limiting bound very restrictive for the possible range within our scan. 
Band structure discussions based on a BdG approach\cite{lederer:2014,bouhon:2014} suggest that for Sr$_2$RuO$_4$ the ratio $K_1/K_2< 1$, which is far from the isotropic limit and even in the regime where current reversal can occur. Furthermore, real sample surfaces are most likely far from specular scattering, but rather belong to the limit of a rough surface\cite{lederer:2014}. While it remains difficult to reliably determine the parameters corresponding to a given sample of Sr$_2$RuO$_4$, we consider the analysis of the experimental results based on the expectations of a specular scattering surface in the isotropic limit as certainly not realistic.

\begin{acknowledgments}
We thank T.~Bzdusek, M.~Fischer, W.~Huang, C.~Hicks, and S.~Yip for many helpful discussions. This study has been financially supported by a grant of the Swiss National Science Foundation.
\end{acknowledgments}

\appendix

\section{The Ginzburg-Landau expansion coefficients}\label{app:para}

In this appendix we provide an overview of the scheme to derive the relation between the different expansion coefficients in the Ginzburg-Landau free energy as given in Eq.~(\ref{eq:bulk}). The chiral $p$-wave state belongs to the two-dimensional irreducible odd-parity representation $ E_u $ of the tetragonal point group $ D_{4h} $. The gap function in d-vector notation is in its most general form given by
\begin{equation}
\boldsymbol{d} (\boldsymbol{k}) = \eta_0 (\phi_x(\boldsymbol k)\eta_x(\boldsymbol{r})+\phi_y(\boldsymbol k)\eta_y(\boldsymbol{r})) \boldsymbol{\hat{z}}
\end{equation}
where $ \{ \phi_x (\boldsymbol k),  \phi_y (\boldsymbol k) \} $ are the two basis functions of $ E_u $. In a weak coupling approach and restricted to a single band (for Sr$_2$RuO$_4$ the $\gamma$-band as most dominant among the three bands), the coefficients of the fourth order terms of the free energy are obtained through
\begin{subequations}
\begin{align}
b_1& = b \langle\phi_x^4\rangle_\mathrm{FS}\\
b_2&= 2 b \langle\phi_x^2\phi_y^2\rangle_\mathrm{FS}\\
b_3&=2(b_2-b_1),
\end{align}
\end{subequations}
where $\langle\cdot\rangle_\mathrm{FS}$ is the average over the Fermi surface and $ b $ is an appropriate parameter depending on specific material properties. For the gradient terms, the same procedure leads to
\begin{subequations}
\begin{align}
K_1& = K \langle v_x^2\phi_x^2\rangle_\mathrm{FS}\\
K_2& = K \langle v_y^2\phi_x^2\rangle_\mathrm{FS}\\
K_3=K_4&= K \langle v_x v_y\phi_x\phi_y \rangle_\mathrm{FS},
\end{align}
\end{subequations}
where $\boldsymbol v$ is the Fermi velocity and $ K $ is a parameter analogous to $ b$ . 

In the isotropic limit, that is, for a rotationally symmetric system with a cylindrical Fermi surface of radius $k_F$, both the momentum and the Fermi velocity reduce to $\boldsymbol k = k_F (\cos\varphi,\sin\varphi)$ and $ \boldsymbol v = v_F (\cos\varphi,\sin\varphi)$, and the basis functions can be chosen as $ \phi_i(\boldsymbol k)= k_i$. Taking the average over the Fermi surface by integration this directly leads to 
\begin{equation}
b_1 = \frac{3}{8} k_F^4 b, \; \;  b_2 = - b_3 = \frac{1}{4} k_F^4 b
\end{equation}
and
\begin{equation}
K_1 = \frac{3}{8} v_F^2 k_F^2 K , \; \; K_2= K_3= K_4 = \frac{1}{8} v_F^2 k_F^2 K.
\end{equation}
This yields the ratios for the isotropic limit, frequently used above,
\begin{equation}
2 b_1 = 3 b_2 = - 3 b_3
\end{equation}
and 
\begin{equation}
K_1 = 3 K_2 = 3 K_3 = 3 K_4 .
\end{equation}
For further details we refer to Refs.~[\onlinecite{etter:2017}] and [\onlinecite{bouhon:2014b}].

\section{Selected analytic expressions}\label{app:fe}

In this appendix, selected analytical expressions derived from the full Ginzburg Landau (GL) free energy functional are collected to enhance the arguments based on the numerical results.

\subsection{Symmetry of the free energy functional}\label{app:symm}

The free energy functional displays the symmetry
\begin{equation}
\mathcal F[\eta_x,\eta_y,D_x,D_y,K_1,K_2]=\mathcal F[\eta_x,\eta_y,D_y,D_x,K_2,K_1]
\end{equation}
for any weak-coupling approach with $K_3=K_4$. This still holds for boundary conditions identical for both order parameter components, i.e. when the effective critical temperature is the same for both components, which includes all surface types with $g_{3,4}=0$ (range C) and the model of an impurity with locally reduced $T_c$. Note that switching $D_x$ and $ D_y$ results in a sign change for the magnetic flux $B_z$, because $\varepsilon_{ijk}=-\varepsilon_{jik}$ in $\nabla\times \boldsymbol A$, such that $B_z(K_1,K_2)=-B_z(K_2,K_1)$. Including the requirements on the surface coefficients $g_i$, this leads to Eq.~(\ref{eq:oddB}), which is confirmed for the surface types C and for an impurity.

\subsection{Expressions derived by variation}\label{app:bc}

For the disk geometry (radius $R$) it is convenient to use polar coordinates $\boldsymbol r=(r,\theta)$ and the polar representation of the order parameter components $\boldsymbol\eta=(\eta_{\perp},\eta_{\parallel}) $ defined in Eq.~(\ref{eq:para}), such that the full free energy is
\begin{equation}
\mathcal{F}\left[\eta_{\perp}(r,\theta),\eta_{\parallel}(r,\theta),A_{\perp}(r,\theta),A_{\parallel}(r,\theta)\right].
\end{equation}
This lengthy expression is omitted here. Rather, the boundary conditions for the order parameters and the full expression for the angular current derived by variation are discussed below.

\subsubsection*{Boundary conditions for the order parameters}

The boundary conditions (BC) from the variation with respect to the order parameter components $\eta_{\perp}$ and $\eta_{\parallel}$ are given below, where the space dependence $(R,\theta)$ is omitted for compactness of the expressions, and where $\widetilde K=(K_1-K_2-2K_3)$.
\begin{widetext}
\begin{subequations}\label{eq:app:bc:full}
\begin{align}
\begin{split}
&-4 (g_1+g_3) \eta_{\perp}= \left(3K_1+K_2+2K_3+\cos(4\theta)\widetilde K\right)		\partial_r\eta_{\perp}\\
&\quad\quad+\left(K_1-K_2+2K_3 - \cos(4\theta) \widetilde K\right)		\left(\frac{\eta_{\perp}}{R}+\frac{\partial_\theta\eta_{\parallel}}{R}\right)-\sin(4\theta)\widetilde K		\left(\frac{\partial_\theta\eta_{\perp}}{R}-\frac{\eta_{\parallel}}{R}+\partial_r\eta_{\parallel}\right) \\
&\quad\quad-i\gamma A_{\parallel} \left(\left(K_1 -K_2 + 2 K_3- \cos(4\theta)\widetilde K\right) \eta_{\parallel}  - \sin(4\theta)\widetilde K \eta_{\perp}\right)\\
&\quad\quad-i\gamma A_{\perp} \left(\left(3 K_1 + K_2 + 2 K_3+  \cos(4\theta)\widetilde K\right) \eta_{\perp} - \sin(4\theta)\widetilde K \eta_{\parallel}\right),
\end{split}\\
\begin{split}
&-4 (g_1-g_3)  \eta_{\parallel}=\left(K_1+3K_2-2K_3-\cos(4\theta)\widetilde K\right)		\partial_r\eta_{\parallel}\\
&\quad\quad-\left(K_1-K_2+2K_3 -\cos(4\theta) \widetilde K\right)		\left(\frac{\eta_{\parallel}}{R}-\frac{\partial_\theta\eta_{\perp}}{R}\right)+\sin(4\theta)\widetilde K		\left(\frac{\partial_\theta\eta_{\parallel}}{R}+\frac{\eta_{\perp}}{R}-\partial_r\eta_{\perp}\right) \\
&\quad\quad-i\gamma A_{\parallel} \left(\left(K_1 -K_2 + 2 K_3- \cos(4\theta)\widetilde K\right) \eta_{\perp}  + \sin(4\theta)\widetilde K \eta_{\parallel}\right)\\
&\quad\quad-i\gamma A_{\perp} \left(\left(K_1 + 3K_2 - 2 K_3-  \cos(4\theta)\widetilde K\right) \eta_{\parallel} - \sin(4\theta)\widetilde K \eta_{\perp}\right).
\end{split}
\end{align}
\end{subequations}
\end{widetext}

Three special cases are discussed further, straight surfaces at the angles $\theta=\{0,\pi/4\}$, the isotropic limit $K_1/K_2=3$, and the specular scattering surface type $s=8$.

The limit of a straight surface is $R\rightarrow\infty$ and $A_{\perp}=0$ for a choice of gauge. At $\theta=0$, Eq.~(\ref{eq:fullbc}) is recovered,
\begin{subequations}
\begin{align}
&-(g_1+g_3)\eta_x=K_1\partial_x\eta_x-i\gamma A_y K_3 \eta_y\\
&- (g_1-g_3)  \eta_y=K_2\partial_x\eta_y-i\gamma A_y K_3\eta_x.
\end{align}
\end{subequations}
For comparison, at the angle $\theta=\pi/4$, the BCs are
\begin{subequations}
\begin{align}
\begin{split}
&-2 (g_1+g_3)  \eta_\perp=\\
&\quad\big(K_1+K_2+2K_3\big)\partial_n\eta_\perp-i\gamma A_\parallel \big(K_1 - K_2\big) \eta_\parallel
\end{split}\\
\begin{split}
&-2 (g_1-g_3)  \eta_\parallel=\\
&\quad\big(K_1+K_2-2K_3\big)\partial_n\eta_\parallel-i\gamma A_\parallel \big(K_1 -K_2\big) \eta_\perp.
\end{split}
\end{align}
\end{subequations}

In the isotropic limit we have $\widetilde K=0$, and for a proper choice of gauge the angular dependence can be written as $ \boldsymbol{\eta}=(\eta_{\perp},\eta_{\parallel})=\eta_{\rm b} (u(r),i v(r))e^{iN\theta}$, with
$A_{\parallel}(r,\theta) = A_{\parallel}(r)$ and $A_{\perp}=0$, such that
\begin{subequations}\label{eq:app:iso}
\begin{align}
&-4 (g_1+g_3)  \eta_{\perp}=3K\partial_r\eta_{\perp}+K\frac{\eta_{\perp}}{R}-i	\left(\gamma A_{\parallel}	-\frac{N}{R}\right)	 K \eta_{\parallel}\\
&-4 (g_1-g_3)  \eta_{\parallel}=K\partial_r\eta_{\parallel}-K\frac{\eta_{\parallel}}{R}-i	\left(\gamma A_{\parallel} 	-\frac{N}{R}\right)	K \eta_{\perp}.
\end{align}
\end{subequations}

For the specular scattering surface type $s=8$, the perpendicular component is fully suppressed at the surface, $\eta_{\perp}(R)=0$. While for straight surfaces (limit $R\to\infty$) the parallel component $\eta_{\parallel}$ has a vanishing slope at the surface, the slope is finite for curved surfaces with finite $R$. At the angle $\theta=0$, the slope is given by
\begin{subequations}
\begin{equation}
\label{eq:app:spec:slopets}
K_2\partial_r \eta_{\parallel}\Big|_R=K_3\frac{\eta_{\parallel}(R)}{R},
\end{equation}
and at the angle $\theta=\pi/4$ by
\begin{equation}
\label{eq:app:spec:slopetd}
(K_1+K_2-2K_3)\partial_r \eta_{\parallel}\Big|_R=(K_1-K_2)\frac{\eta_{\parallel}(R)}{R},
\end{equation}
\end{subequations}
where we have used the fact that symmetry arguments lead straightforwardly to $A_\perp(\theta=0)=A_\perp(\theta=\pi/4)=0$ for our choice of gauge, as described in Ref.~[\onlinecite{etter:2017}].
In the isotropic limit as described above, this results in a positive slope at all angles (see Fig.~\hyperref[fig:iso]{\ref*{fig:iso}(a)}).
Away from the isotropic limit, the slope at $\theta=0$ always remains positive, while the slope at $\theta=\pi/4$ vanishes when $K_1=K_2$ and changes sign for $K_1<K_2$.
We further note that these equations even hold for concave surfaces, but that due to the negative curvature the slope has opposite sign at the surface, leading for example to a slight reduction of $\eta_\parallel$ at the surface in the isotropic limit.
In any case, the initial enhancement of the parallel order parameter due to the suppression of the perpendicular component always remains and dominates the overall behavior.



\subsubsection*{Full expression for the angular current}

The full GL equation for the parallel current $J_{\parallel}(r,\theta)$ (not restricted to the surface) derived by variation  is given below, where the space dependence $(r,\theta)$ is again omitted for compactness of the expressions, and where again $\widetilde K=(K_1-K_2-2K_3)$.

\begin{widetext}
\begin{equation}\label{eq:app:curr:full}
\begin{split}
-\frac{2}{c\gamma}J_{\parallel}=
&\left(K_1-K_2+2K_3\right)\Im\left[\eta_{\perp}\partial_r\eta_{\parallel}^*-\partial_r\eta_{\perp}\eta_{\parallel}^*\right]\\
&-\frac{1}{r}\Im\left[\left(4\left(K_1+K_2\right)\eta_{\perp}\eta_{\parallel}^*-\left(K_1+3K_2-2K_3\right)\eta_{\perp}\partial_\theta\eta_{\perp}^*-\left(3K_1+K_2+2K_3\right)\eta_{\parallel}\partial_\theta\eta_{\parallel}^*\right)\right]\\
&-\widetilde K\Im\left[\eta_{\perp}\partial_r\eta_{\parallel}^*-\partial_r\eta_{\perp}\eta_{\parallel}^*+\frac{1}{r}\left(\eta_{\perp}\partial_\theta\eta_{\perp}^*-\eta_{\parallel}\partial_\theta\eta_{\parallel}^*\right)\right]\cos(4\theta)\\
&-\widetilde K \Im\left[\eta_{\perp}\partial_r\eta_{\perp}^*-\eta_{\parallel}\partial_r\eta_{\parallel}^*-\frac{1}{r}\left(\eta_{\perp}\partial_\theta\eta_{\parallel}^*-\partial_\theta\eta_{\perp}\eta_{\parallel}^*\right)\right]\sin(4\theta)\\
&+\gamma A_{\perp}\left(\left(K_1-K_2+2K_3-\cos(4\theta)\widetilde K\right)2\Re\left[\eta_{\perp}\eta_{\parallel}^*\right]-\sin(4\theta)\widetilde K\left(|\eta_{\perp}|^2-|\eta_{\parallel}|^2\right)\right)\\
&+\gamma A_{\parallel}\Big(
2\left(K_1+K_2\right)|\boldsymbol\eta|^2-\left(K_1-K_2+2K_3\right)\left(|\eta_{\perp}|^2-|\eta_{\parallel}|^2\right)\\
&\quad\quad\quad\,\,-\cos(4\theta)\widetilde K\left(|\eta_{\perp}|^2-|\eta_{\parallel}|^2\right)+\sin(4\theta)\widetilde K2\Re\left[\eta_{\perp}\eta_{\parallel}^*\right]\Big).
\end{split}
\end{equation}
\end{widetext}

Three special cases are discussed further,  the isotropic limit $K_1/K_2=3$, the specular scattering surface type $s=8$, and full pair breaking $s=18$.

Treating the isotropic limit as above for Eq.~(\ref{eq:app:iso}), the `edge' and the `screening' part of the total angular current $ J_{\parallel} $ are extracted by demanding that both parts vanish in the bulk, leading to
\begin{subequations}\label{app:eq:jparts}
\begin{align}
\frac{2}{c\gamma K}J_\mathrm{edge}&= \Im\left[\partial_r\eta_{\perp}\eta_{\parallel}^*-\eta_{\perp}\partial_r\eta_{\parallel}^*\right]\\
\frac{2}{c\gamma K}J_\mathrm{screen}&=\frac{4}{r}\Im\left[\eta_{\perp}\eta_{\parallel}^*\right]-\left(\gamma A_{\parallel}-\frac{N}{r}\right) \left(|\eta_{\perp}|^2+3 |\eta_{\parallel}|^2\right).
\end{align}
\end{subequations}

For specular scattering with $\eta_{\perp}(R)=0$ ($s=8$) and for any ratio $K_1/K_2$ (not restricted to the isotropic limit), the current at the surface $r=R$ and at the angle $\theta=0$ is given by
\begin{subequations}\label{app:eq:jspec}
\begin{equation}
\frac{J_{\parallel}(R)}{2c\gamma}=\Im\left[K_3\partial_r\eta_{\perp} \eta_{\parallel}^* -K_1\eta_{\parallel}\frac{ \partial_\theta\eta_{\parallel}^* }{R}\right]-   \gamma A_{\parallel} K_1 |\eta_{\parallel}|^2,
\end{equation}
and at the angle $\theta=\pi/4$ by
\begin{equation}\label{eq:jtspecp4}
\begin{split}
\frac{J_{\parallel}(R)}{c\gamma}=&\Im\left[\left(K_1-K_2\right)\partial_r\eta_{\perp} \eta_{\parallel}^*-\widehat K\eta_{\parallel} \frac{\partial_\theta\eta_{\parallel}^*}{R}\right]
-\gamma A_{\parallel} \widehat K |\eta_{\parallel}|^2,
\end{split}
\end{equation}
\end{subequations}
where $\widehat K=(K_1 + K_2 + 2 K_3)$.
For a straight surface with $R\rightarrow\infty$, the driving term $J_\mathrm{edge}$ of Eq.~(\ref{eq:jtspecp4}) vanishes for $K_1=K_2$. Therefore, the screening term also vanishes and there is current reversal for any $K_1<K_2$.

For full pair breaking with $\eta_{\perp}(R)=0$ and $\eta_{\parallel}(R)=0$ ($s=18$) and for all ratios $K_1/K_2$, the angular current at the surface $r=R$ vanishes,
\begin{equation}\label{app:sub:fpb}
J_{\parallel}(R)=0.
\end{equation}

\bibliography{flux_ref}

\end{document}